\documentclass[a4paper,12pt]{article}   
  \usepackage{makeidx}
  \usepackage{graphicx,graphics,color}
  \usepackage{latexsym,doc,fontenc,exscale}
 \usepackage{amsmath,amsfonts,amsthm,eucal,amssymb}

 \DeclareMathOperator{\sech}{sech}

\begin{document}
 \title{\bf  Tracking down localized modes in PT-symmetric Hamiltonians under the influence of a competing nonlinearity}
 \author{\textbf{Bijan Bagchi}\\
 \small{Department of Applied Mathematics, University of Calcutta},\\ \small{92 Acharya Prafulla Chandra Road, Kolkata-700 009, India}\\
 \small{bbagchi123@rediffmail.com}\\\\
 \textbf{Subhrajit Modak and Prasanta K. Panigrahi}\\
 \small{Department of Physical Science,}\\
 \small{Indian Institute of Science, Education and Research (Kolkata),  }\\
 \small{Mohanpur, West Bengal 741 252, India}\\
 \small{modoksuvrojit@gmail.com, panigrahi.iiser@gmail.com}}

\date{ }
 \maketitle
 \noindent


 The relevance of parity and time reversal (PT)-symmetric structures in optical systems is known for sometime with the correspondence existing between the Schr$\ddot{o}$dinger equation and the paraxial equation of diffraction where the time parameter represents the propagating distance and the refractive index acts as the complex potential. In this paper, we systematically analyze a normalized form of the nonlinear Schr$\ddot{o}$dinger system with two new families of PT-symmetric potentials in the presence of competing nonlinearities. We generate a class of localized eigenmodes and carry out a linear stability analysis on the solutions. In particular, we find an interesting feature of bifurcation charaterized by the parameter of perturbative growth rate passing through zero where a transition to imaginary eigenvalues occurs.

 \newpage

 \section{Introduction}

 Following Bender and Boettcher's seminal paper \cite{bend1} in which they offered the first coherent explanation of a special class of non-Hermitian but parity and time-reversal (PT)-symmetric Hamiltonians to possess a real bound-state spectrum, the last decade has witnessed extensive theoretical work \cite{bend2,bend3,mosta} being devoted to this
 growing field of research. The interplay between the parametric regions where PT is unbroken and the ones in which PT is broken as signaled by the appearance of conjugate-complex eigenvalues (see, for example, \cite{zn,bq,ap}) has for sometime found repeated experimental support \cite{guo,ruter,rubi,zhao,chong,lin,zheng,bittner,schindler,zameit} as evidenced by the observations of a phase transition that clearly marks out the separation of these regions. It is useful to bear in mind that the analytical studies in this regard have mostly been carried out for the linear domain. Of late, the relevance of PT-structure has been noticed in various optical systems and interesting features seen such as, for example, the power oscillations \cite{ruter}, unidirectional invisibilty \cite{lin}, coherent perfect absorber \cite{longhi, chong}, giant wave amplification \cite{kon} and realiztion through electromagnetically induced transparency \cite{li}. In optical systems, PT-symmetry has the implication that the index guiding part $n_{R}(x)$ and the gain/loss profile $n_{I}(x)$ of the complex refractive index  $n(x) = n_{R}(x) + in_{I}(x)$ obey the symmetric $n_{R}(x)=n_{R}(-x)$ and antisymmetric $n_{I}(x)=-n_{I}(-x)$ combinations (see, for example, \cite{kla,ben,reg}). Balancing gain and loss \cite{musli1, musli2, musli3,musli4} is an interesting curiosity towards experimental realization of PT-symmetric Hamiltonians.

 Against the background of the experimental findings, Musslimani et al \cite{musli1,musli2} have reported optical solitons in PT-periodic potentials which are stable over a wide range of potential parameters. Specifically they have considered optical wave propagation with the beam evolution being controlled by a normalized nonlinear Schr$\ddot{o}$dinger (NLS) equation defined in terms of an electric field envelop and a scaled propagation distance. Indeed, the generalized NLS they consider, in the presence of a PT-symmetric potential, is given by
 \begin{equation}\label{33}
 i\psi_{z}+\psi_{xx}+[V(x)+iW(x)]\psi+g|\psi|^{2}\psi=0
 \end{equation}
 with the PT-symmetric potential possessing the usual properties \cite{br} $V(-x)=V(x)$ and $W(-x)=-W(x)$. In (1) $\psi$ represents the electric field envelope, $z$ is a scaled propagation distance and $g=1$ or $-1$ corresponds to a self-focussing or a defocussing nonlinearity. Further, the $\psi_{xx}$  term describes the optical diffraction, $V(x)$ is the index guiding and $W(x)$ represents the gain/loss distribution of the optical potential. Musslimani et al \cite{musli1,musli2} studied nonlinear stationary solutions of the form $\psi(x,z)=
 \phi(x)\exp(i\lambda z)$, $\lambda$ being a real propagation constant and $\phi$ is the signature of the nonlinear eigenmode. In the context of nonlinear optics, localised modes are either temporal or spatial depending on whether the confinement of light occurs in time or space during wave propagation. Temporal modes correspond to optical pulses that maintain their shapes whereas spatial modes represent propagating transverse self-guided beams orthogonal to the direction of movement. These modes are termed as solitons. Both the types of solitons emerge from a nonlinear change in the refractive index of an optical material induced by the light intensity. This phenomenon is referred to as the optical Kerr effect. The intensity dependence of the refractive index leads to spatial self-focussing (or self-defocussing) and temporal self-phase modulation, the two major nonlinear effects that are responsible for the formation of optical solitary modes or optical solitons \cite{agg}.

 In this article we report on some new localized solutions of the NLS and study the distribution of eigenmodes on the real and complex plane by incorporating the effects of higher degree nonlinear effects over and above the minimal cubic term. By parametrizing the coupling strength of the latter and arbitrarily specifying the order of additional nonlinearity on a Rosen-Morse potential we observe numerically for one class of solutions the existence of a threshold value of the growth rate parameter beyond which suitably chosen pair of discrete eigenmodes on the real axis merge and subsequently appear in conjugate imaginary pairs exhibiting the qualitative character of bifurcation. In this connection it needs to be pointed out that our model differs significantly from those advanced so far to search for solitonic solutions \cite{ka,mi}. For instance, the potentials of our interest are markedly different from the Rosen-Morse type considered in \cite{mi} because of the presence of an additional nonlinear term in our case. The PT-symmetric potentials addressed in \cite{ka} are basically nonlinear extensions of the Scarf II. Also, the above aspect of bifurcation did not arise in the models considered in \cite{ka,mi}.

 \section{Mathematical Model and Formulation}
\ We are considering an optical wave propagation in the presence of a  $\mathcal{PT}$-symmetric potential. In this case the beam dynamics is governed by a generalized nonlinear Schr\"odinger model with competing nonlinearities, i.e.,
\begin{equation}
i \frac{\partial \Psi}{\partial z} + \frac{\partial^2 \Psi}{\partial x^2} + \left[ V(x)+ i W(x)\right] \Psi + g_{1} |\Psi|^2 \Psi+g_{2}|\Psi|^{2\kappa}\Psi =0.{\label{e1}}
\end{equation}
where $\kappa$ is an arbitrary real number, $\Psi(x,z)$ is complex electric field envelope, $g_{1}$ and $g_{2}$ control respectively the strength of the cubic and arbitrary nonlinear term. It is clear that Eq.(2) admits stationary solutions  $\Psi(x,z) = \phi(x) e^{i \lambda z}$, where $\lambda$ is a real propagation constant and the complex function $\Phi(x)$ obeys the eigenvalue equation
\begin{equation}
\frac{\partial^2 \Phi}{\partial x^2}+\left[ V(x)+ i W(x)\right] \Phi+g_{1} |\Phi|^2\Phi+g_{2}|\Phi|^{2\kappa}\Phi=\lambda\Phi.
\end{equation}
We now show that this model supports two different soliton solutions marked by Class I and Class II cases provided we do not alter the imaginary part of the potential but only choose the real part appropriately.

\subsection{Class I solutions}
We focus on a  $\mathcal{PT}$-symmetric Rosen-Morse potential  $-a (a + 1) \sech^2 x +  2i b \tanh x ~(a,~b\in\mathbb{R})$ being subjected to an additional term
$-V_{1} \sech^{2\kappa} x~~(V_{1},~\kappa\in\mathbb{R})$, i.e.,
\begin{equation}
 V(x) = -a (a + 1) \sech^2 x-V_{1} \sech^{2\kappa} x,~
 W(x) =  2 b \tanh x.
\end{equation}
Corresponding to $(4)$ there always exists for $(3)$ a typical solution
\begin{equation}
\Phi(x)=\Phi_{0} \sech(x) ~e^{i \mu x}
\end{equation}
provided that the amplitude $\Phi_{0}$ and the phase factor $\mu$ are related to the potential parameters through
\begin{equation}
\Phi_{0}^ {2\kappa}=\frac{V_{1}}{g_2},~~\Phi_{0}^2=\frac{a^2+a+2}{g_1},~~
b=\mu,~\lambda=1-\mu^2
\end{equation}
Note that the imaginary strength of the potential contributes only through the phase factor and the amplitude from the real part of the potential are intertwined through the  strength of the nonlinearity irrespective of its sign. For this solution the transverse power flow defined by
\begin{equation}
S = \frac{i}{2} (\phi \phi_x^* -\phi^* \phi_x)
\end{equation}
turns out to be $S=b\Phi_{0}^2{\sech^2(x)}$ implying that the transmission depends upon the strength of the imaginary part of the potential.
\subsection{Class II solutions}
On the other hand, if we choose the extended Rosen-Morse potential to have the form
\begin{equation}
 V(x) = -a (a + 1) \sech^2 x-V_{1} \sech^{\frac{2}{\kappa}} x,~~~
 W(x) =  2 b \tanh x,~~~\kappa\in\mathbb{R} - \{0\}
\end{equation}
then Eq.(2) enjoys a solution
\begin{equation}
\Phi(x)=\Phi_{0} \sech^{\frac{1}{\kappa}} x ~e^{i \mu x}
\end{equation}
if the amplitude and phase factor are constrained by the relations
\begin{equation}
\Phi_{0}^2=\frac{V_{1}}{g_1},~b=\frac{\mu}{\kappa},~\lambda=\frac{1}{\kappa^2}-\mu^2,~\Phi_{0}^{2\kappa}=\frac{1}{g_2}[ a(a+1)+(\frac{1}{\kappa}+\frac{1}{\kappa^2})]
\end{equation}
Note that the solution (10) is valid irrespective of the signs of $g_{1}$ and $g_{2}$. The transverse power flow, defined by Eq.(5), turns out to be
$S=b\Phi_{0}^2 \kappa{\sech^{\frac{2}{\kappa}} x}$ which in this case is influenced by both $b$ and $\kappa$ including of course the effects of their signs.

\section{Numerical computations and Eigenmode distribution}
Solitary waves associated with the non-Kerr nonlinear media retain their shape but their stability is not guaranteed because of the nonintegrable nature of the underlying extended NLS equation we have at hand. In fact, their stability against small perturbation is an important issue beacuse only stable (or weakly unstable) self-trapped beams can be observed experimentally. Let us impose a small perturbation to determine  whether it is stable or unstable against this slight disturbance.
More specifically we consider a perturbation of the form \cite{ka}
\begin{equation}
 \Psi(x,z) = \phi(x) e^{i \lambda z} + \left\{[ v(x) + \omega(x) ] ~e^{\eta z} + [v^*(x) -\omega^*(x)]~ e^{\eta^* z}\right\}  e^{i \lambda z}\label{e30}
\end{equation}
where $v(x)$ and $\omega(x)$ are infinitesimal perturbed eigenfunctions such that $|v|,|\omega| \ll |\phi|$ and $\eta$ indicates the perturbed growth rate. Linearization of Eq.(1) around  $\Phi(x)$ yields the following eigenvalue problem
\begin{equation}
 \left( \begin{array}{cc}
0 & ~\hat{\mathcal{L}}_0  \\
 \hat{\mathcal{L}}_1 & ~0 \\
\end{array} \right)   ~~~ \left( \begin{array}{c}
v   \\
\omega \\
 \end{array} \right)= - i \eta
 \left( \begin{array}{c}
v \\
\omega \\
 \end{array} \right)\label{e12}
\end{equation}
where $\hat{\mathcal{L}}_0 = \partial_{xx} - \lambda+ (V+i W) + g_{1} |\phi|^2+g_{2}|\Phi|^{2\kappa} $ and
$\hat{\mathcal{L}}_1 = \partial_{xx}- \lambda + (V+i W) + 3 g_{1} |\phi|^2+g_2 (1+2\kappa)|\Phi|^{2\kappa} $
 and $\eta$ is the associated eigenvalue corresponds to the growth rate parameter. The $\eta$-spectrum is called the linear-stability spectrum for the localized modes. It is easy to see if $\eta$ is an eigenvalue then so are $\eta^{*}$, -$\eta$, and -$\eta^{*}$ indicating that these eigenvalues always appear in pairs or quadruples.\\

 The continuous spectrum of Eq.(12) can be readily recovered in the large-distance limit of $|x|\rightarrow$~$\infty$. Under this limit, $\hat{\mathcal{L}}_0$ and $\hat{\mathcal{L}}_1$ move over to a simple differential operator with constant coefficient. In order for $\eta$ to be in the continuous spectrum, the corresponding eigenfunction at large distance must be a Fourier mode. If we observe the orientation of the eigenmodes in the entire spectrum, we run into three different kinds of modes. The appearance of nonzero discrete eigenvalues in the linearization spectrum of solitary waves is a signature of nonintegrable character of the equation. If the spectrum contains a real positive eigenvalue, the corresponding eigenmode in the perturbed solution will grow exponentially with time; hence the solitary wave is linearly unstable. Generally if the spectrum contains any eigenvalue with real positive part then such eigenvalues are unstable. Secondly if the spectrum admits of a pair of conjugate-complex eigenvalues (internal modes) the perturbed solution will exhibit oscillations leading eventually to shape fluctuations that would be smothered with time. Thirdly one can encounter zero eigenvalues which are the so-called Goldstone modes (see for a discussion on this point \cite{ka}). The behavior of the eigenvalues $\eta$  can be ascertained by solving Eq.(12) numerically. Here we adopt the Fourier collocation method \cite{ya1, ya2} to track the tendencies of the eigenvalues. We now turn to some discussions of our results.

\section{Results and Discussion}

 Figures $1,~2$,~$3$ and $4$ give a graphical display of our numerical results on the eigenmode distribution. The interplay between the cubic and competing nonlinearity on the soliton dynamics is best understood in terms of the parameters $\Phi_{0}$, $|g_1|$, $|g_2|$. We also look for stability around some specific value of $\kappa$ as mentioned in the Figure captions. It should be borne in mind that Eq.(6) and Eq.(10) constrain these parameters in terms of the amplitude $\Phi_{0}$ and the phase factor $\mu$  as well as the coupling constants of the potential. The evolution of Class I solution for different choices of $g_1$ and $g_2$ is laid down in Fig.1. Here the discrete modes initially lie on the real axis corresponding to a sample choice of $a$, $b$ and $g_2$ - for specific values of these parameters see the corresponding Figure captions. Normalizing $\Phi_{0}$ to unity without any loss of information, $g_1$ gets automatically fixed while $b$, $\lambda$ and $\mu$ acquire their values from the consistency conditions (6). In this manner of parametrization $g_1$ and $g_2$ differ in sign while the magnitude of $g_1$ turns out to be weaker than $g_2$. In Fig.2 the plots are sequentially arranged according to the varying strengths of the couplings corresponding to the cubic nonlinear term and holding  $g_2$ fixed. We note that $g_1$ changes according to the potential parameter $a$. A new type of solution develops at this point due to the sensitive dependence of the perturbative growth rate parameter $\eta$ on $a$: around $a= 0.03$ we see that as the parameter $a$ is varied the discrete modes initially lying on the real axis mutually approach towards the zero-mode eigenvalue. However, further change in $a$ causes a pair of imaginary eigenvalues to develop revealing a typical feature of bifurcation. We next carry out computations for equal and opposite values of $|g_1|$ and $|g_2|$ couplings. In Fig.3 we see that in such a case the real eigenmodes lead to the solitonic solutions becoming unstable. A similar situation exists in Fig.4 where oscillatory instability along the imaginary axis is caused by the equal-strength coupling parameters from the two nonlinear terms.\\
 
 Finally, let us point out that Class II solutions can be evaluated under various parametric conditions. Here we inevitably run into the unstable character of the soliton solutions (in Figure 4 one such situation is described to compare with the counterpart of Class I solution). It should be emphasized that for the various run of the parameters we were unable to track down any feature of bifurcation characteized by the growth rate parameter crossing the zero-threshold value and transiting to the complex plane. \\
 
 We thank Dr. Abhijit Banerjee for useful discussions.


\begin{figure}
\centering

\includegraphics[width=4 cm,height=4.3 cm]{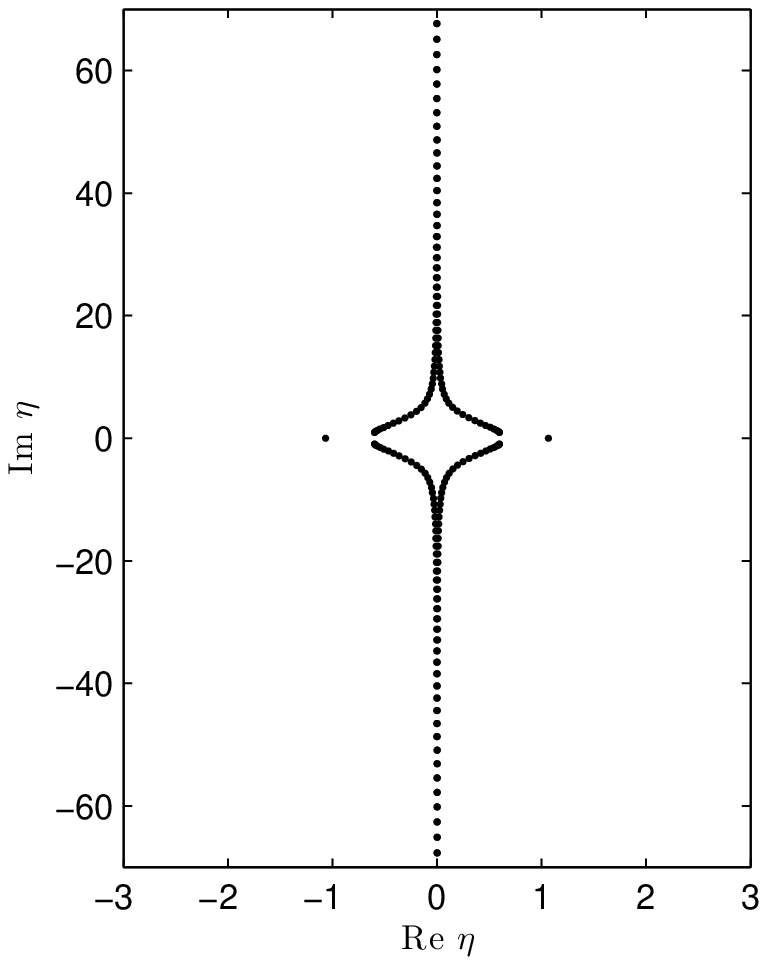}
\caption{Numerically computed eigenmode distribution. In this case we have considered $a=.01$, $b=.3$, $V_{1}=g_{2}=-4$, $g_{1}=2.01$ and $\kappa = 3$. Specification of $g_{1}$ is done by choosing the potential parameter $a$. }

\includegraphics[width=4 cm,height=4.3 cm]{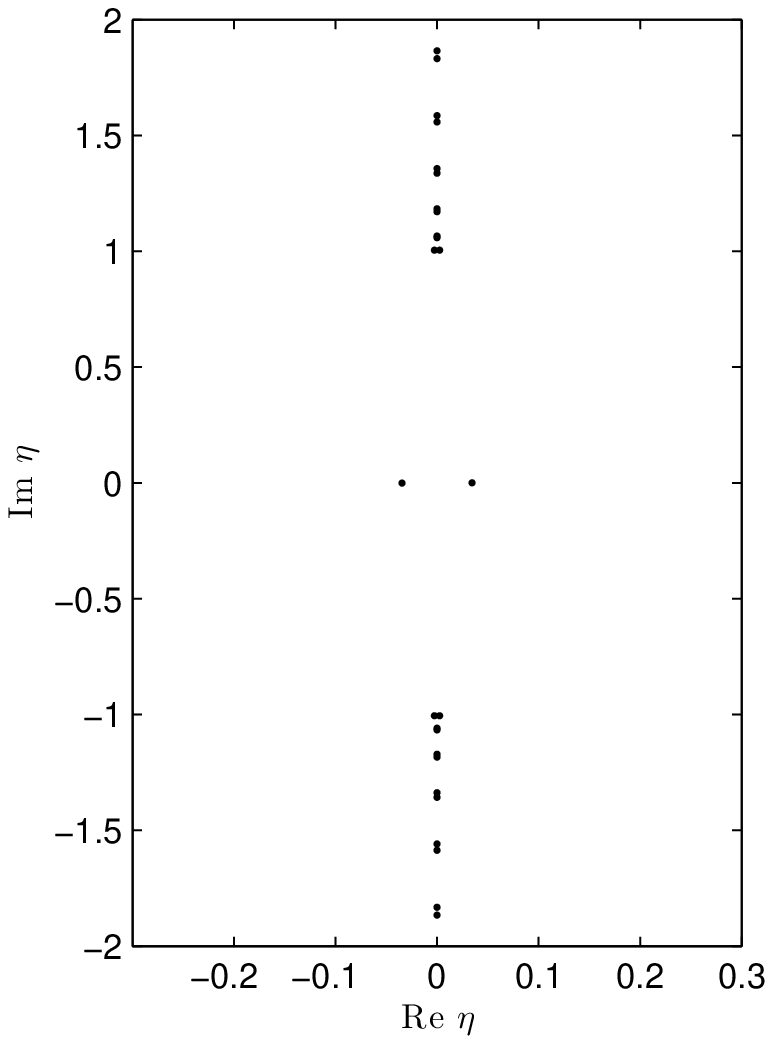}~~~~~\includegraphics[width=4 cm,height=4.3 cm]{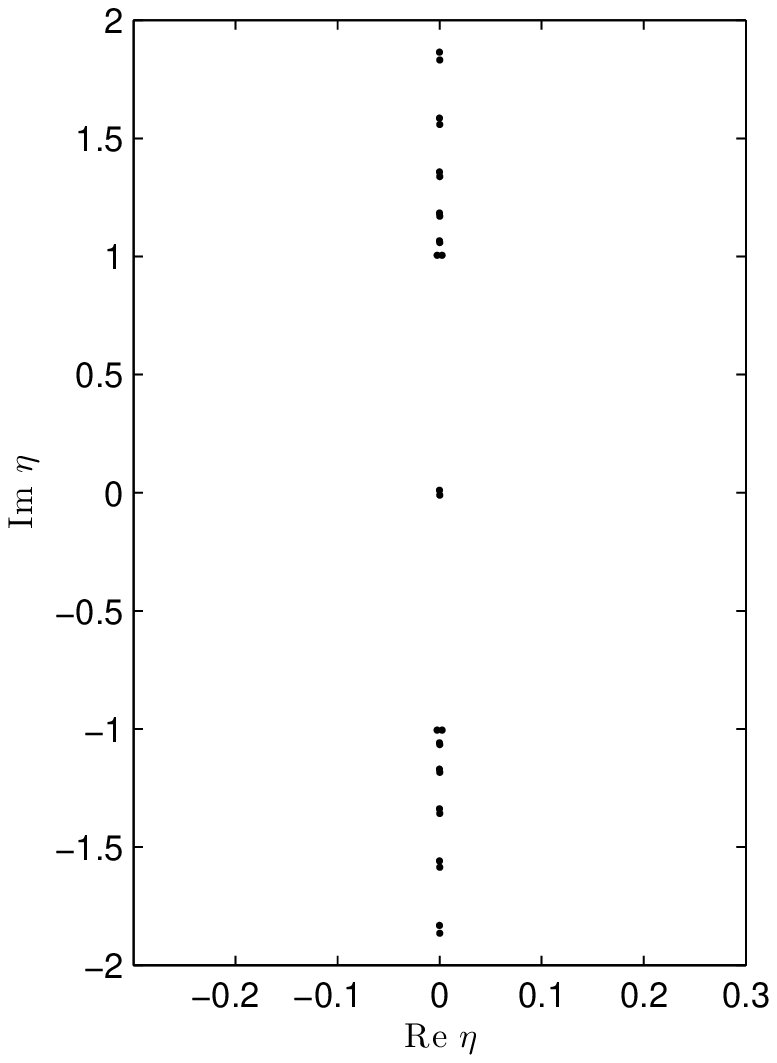}~~~~\includegraphics[width=4 cm,height=4.3 cm]{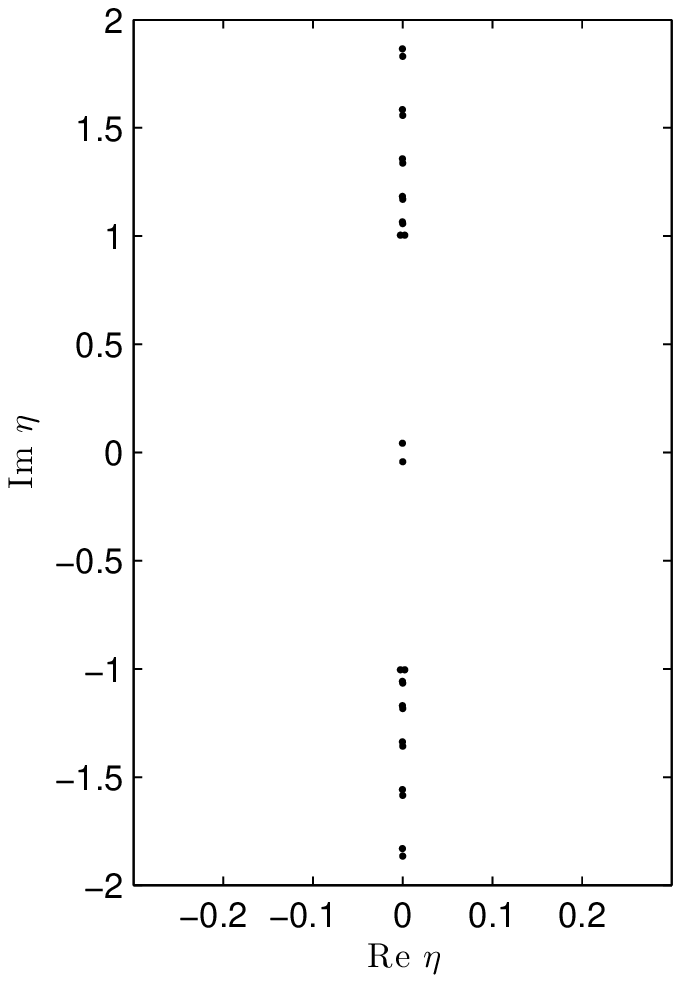}~~~~\includegraphics[width=4 cm,height=4.3 cm]{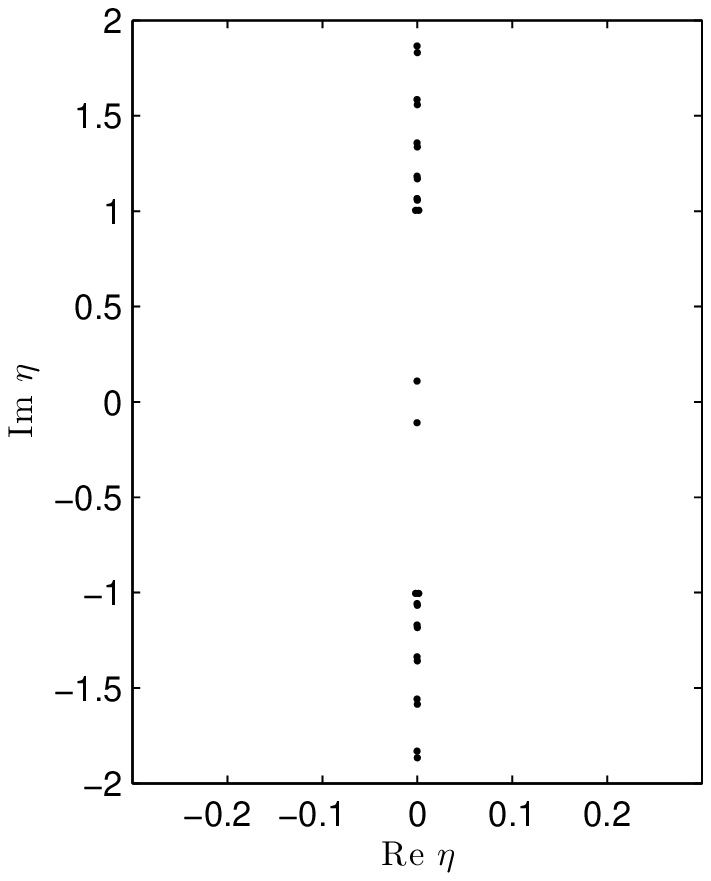}\\

\caption{Sequentially computed eigenmode behavior for $b$=.003, $V_{1}=g_{2}=-4$ and $\kappa = 3$. In this case the coupling parameters corresponding to cubic nonlinearity are varied against the potential parameter $a$ continuously from $a$ =.03 to .09 . Four different values of $a$ as $a$=.03, $a$=.04, $a$=.05, $a$= .09 are considered and the Figures are arranged in such a way that the lowest value of $a$ corresponds to the figure at the left and the highest to the figure at the right.}

\includegraphics[width=4 cm,height=4.3 cm]{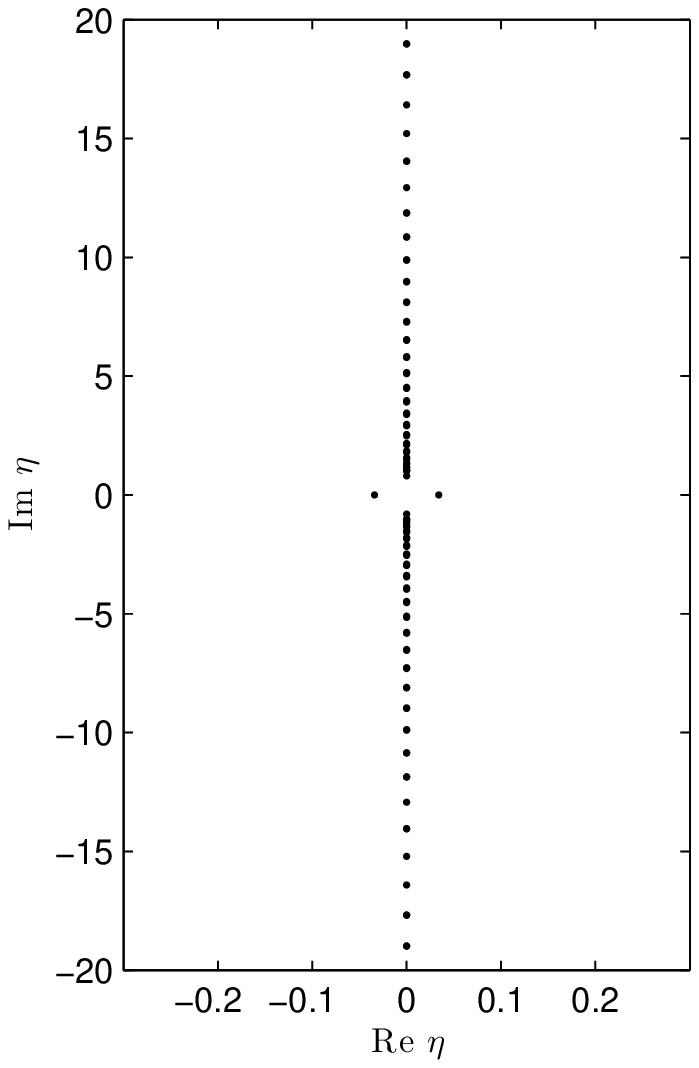}
\caption{(Color online) The figure shows the unstable modes for the coupling parameters $|g_1|$ and $|g_2|$ with equal and opposite strengths.  Distribution of eigenmodes for $a$=1, $b$=.003, $g_{1}$=4, $V_{1}=g_{2}=-4$ and $\kappa = 3$.}
\end{figure}

\begin{figure}
\centering

\includegraphics[width=4 cm,height=4 cm]{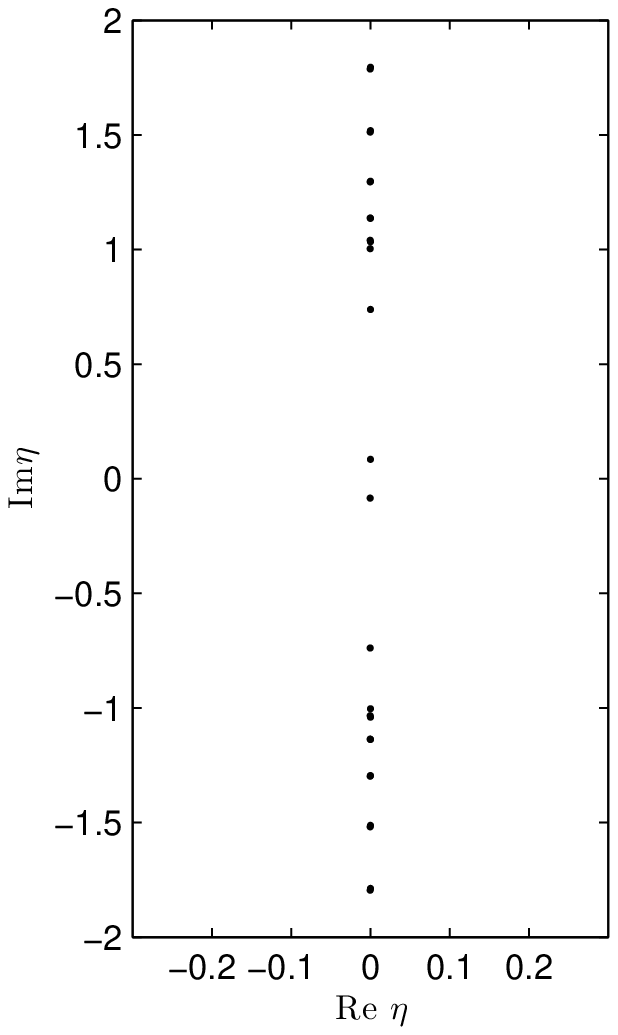}~~~~\includegraphics[width=4 cm,height=4 cm]{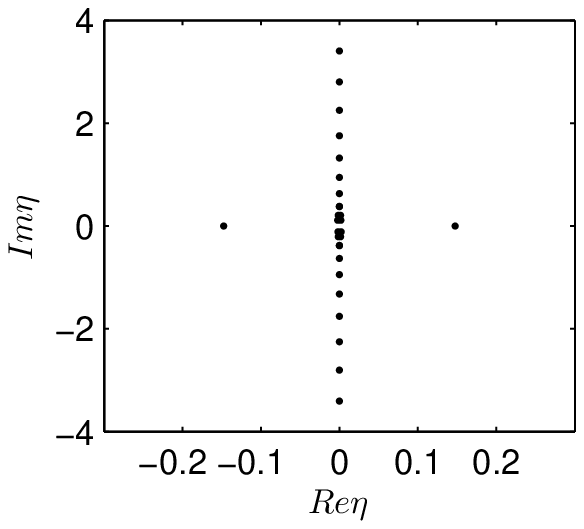}~~~\includegraphics[width=4 cm,height=4 cm]{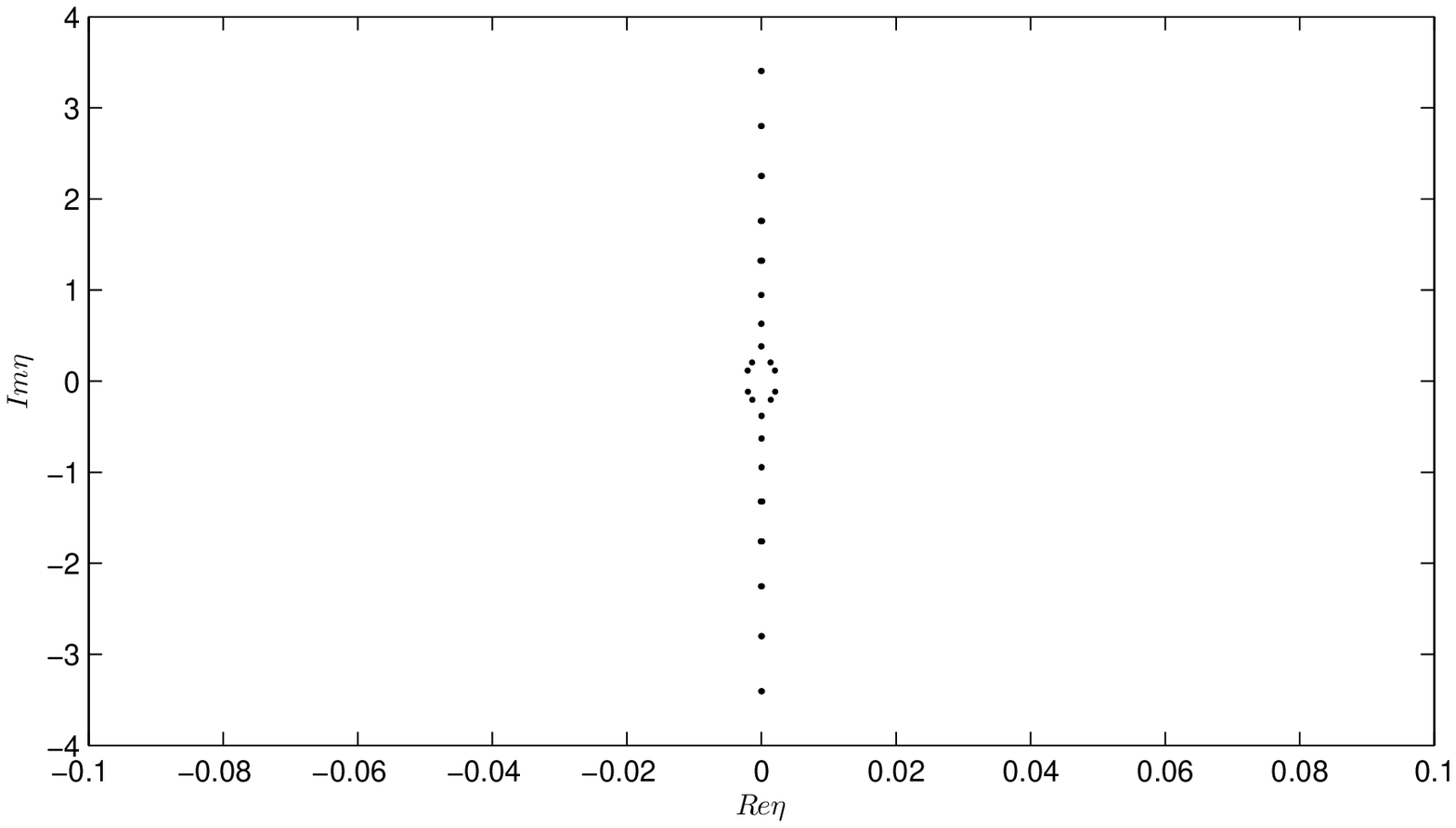}

\caption{Figure at the leftmost showing the unstable modes corresponding to equal-strength couplings for the choice $g_1=4$, $g_2=4$ obtained from (6) and $\kappa = 3$.  The potential parameters are taken to be $a=1$ and $b=.003$ . The figure at the middle corresponds to the evaluation of Class II solution under the same parametrizations. The coupling strengths for Class II are $g_1=4$ and $g_2=2.44$ obtained from (10). The figure at the right is an enlarged description of the behavior near the origin.}

\end{figure}

 \end{document}